\documentclass{article}
\usepackage{geometry}
\newenvironment{Figure}
 {\par\medskip\noindent\minipage{\linewidth}}
 {\endminipage\par\medskip}

\usepackage{multicol, caption}
\usepackage[utf8]{inputenc}
\usepackage{float}
\usepackage{amsmath}
\usepackage{xcolor}
\usepackage{graphicx}
\usepackage[sorting=none, style=chem-rsc, maxbibnames=3]{biblatex}
\usepackage{braket}

\usepackage[colorinlistoftodos]{todonotes}
\usepackage{comment}
\usepackage[super]{nth}
\usepackage{verbatim}
\usepackage{indentfirst}
\usepackage{authblk}
\title{Quantum Computational Quantification of Protein-Ligand Interactions}
\author[1]{Josh J. M. Kirsopp\thanks{josh.kirsopp@cambridgequantum.com}}
\author[1]{Cono Di Paola}
\author[1]{David Zsolt Manrique}
\author[1]{Michal Krompiec}
\author[1]{Gabriel Greene-Diniz}
\author[2]{Wolfgang Guba\thanks{wolfgang.guba@roche.com}}
\author[2]{Agnes Meyder}
\author[2]{Detlef Wolf}
\author[2]{Martin Strahm}
\author[1]{David Mu{\~n}oz Ramo}
\affil[1]{\textit{Cambridge Quantum Computing Ltd, 13-15 Hills Road, CB2 1NL, Cambridge, UK}}
\affil[2]{\textit{Roche Pharmaceutical Research \& Early Development, Roche Innovation Center Basel, F. Hoffmann-La Roche Ltd., Basel, 4070, Switzerland}}
\date{15 October 2021}

\begin{document}

\maketitle

\begin{abstract}\label{section:abstract}
We have demonstrated a prototypical hybrid classical and quantum computational workflow for the quantification of protein-ligand interactions. The workflow combines the Density Matrix Embedding Theory (DMET) embedding procedure with the Variational Quantum Eigensolver (VQE) approach for finding molecular electronic ground states. A series of $\beta$-secretase (BACE1) inhibitors is rank-ordered using binding energy differences calculated on the latest superconducting transmon (IBM) and trapped-ion (Honeywell) Noisy Intermediate Scale Quantum (NISQ) devices. This is the first application of real quantum computers to the calculation of protein-ligand binding energies. The results shed light on hardware and software requirements which would enable the application of NISQ algorithms in drug design. 
\end{abstract}

\begin{multicols}{2}

\section{Introduction}
The advent of quantum mechanics at the turn of the 20th century changed the way we look at the physical sciences. For chemistry, the implications were profound, as Dirac famously noted: ``the fundamental laws ... for the whole of chemistry are thus completely known, and the difficulty lies only in the fact that application of these laws leads to equations that are too complex to be solved'' \cite{DiracQuote}. Indeed, calculations of accurate solutions of the electronic Schr\"{o}dinger equation, such as the Full Configuration Interaction method (FCI), of molecules scale exponentially with the number of atoms \cite{helgaker2014molecular}, rendering them applicable only to the smallest systems. Practical approximations to FCI, such as e.g. CCSD(T) \cite{RevModPhys.79.291}, touted as computational chemistry's gold standard, do exist, but their applicability is limited: single-reference methods like CCSD(T) fail for strongly-correlated (``multireference'') systems and the formal scaling of CCSD(T) with system size is O($N^7$), rendering it useful only for relatively small molecules. While the scaling limitation of coupled cluster methods (CC) can be somewhat overcome via localized methods such as DLPNO \cite{NeeseDLPNOCCSDT} at the expense of accuracy, methods for strongly-correlated systems scale exponentially with the number of correlated electrons and, even for systems small enough to be tractable, require expert knowledge to be applied. On the other hand, mean-field methods, such as Hartree-Fock (HF) and Density Functional Theory (DFT), which either dispense with electron correlation completely or treat it in an approximate, implicit manner, are routinely applied to organic and inorganic systems of sizes up to few hundred atoms. This is possible due to their relatively low, $O(N^3)$ - $O(N^4)$, formal scaling, which could be reduced even to $O(N)$ in approximate implementations for systems comprised of thousands of atoms \cite{ONETEP,BIGDFT,linear_scaling_rev_2013}. However, HF and DFT fail for multi-reference (strongly-correlated) systems and cannot describe dispersion interactions, which are a critical component of inter-molecular forces, otherwise than via an ad hoc correction \cite{Grimme_review_2016}.  Clearly, general-purpose, low-scaling and highly accurate electronic structure methods applicable to arbitrary molecular systems remain elusive.  
It is widely believed that the simulation of complex and strongly-correlated chemical systems, for which standard methods are either inaccurate or too expensive, is among the areas which stand to benefit most from the ongoing and rapid progress in quantum computation \cite{Troyer_Reiher_2021_rev}. Indeed, the last half-decade has seen an eruption of research in quantum algorithms for materials simulations, including computation of ground and excited states of molecules, quantum dynamics and linear response, among a plethora of others \cite{bharti2021noisy, mcardle2020quantum, lanyon2010towards, cao2019quantum, bauer2020quantum}. Noisy Intermediate Scale Quantum (NISQ) devices, however, have so far limited the applicability of these algorithms to very small systems, such as H$_2$, LiH, RbH and the like \cite{kubra2021,doi:10.1063/5.0044068}. Nevertheless, rapid progress in the capabilities of quantum hardware as well as intensive research on new quantum algorithms opens the possibility of utilising quantum computing in Computer Aided Drug Design (CADD) in the future. 

The synthesis of new drug candidates requires substantial medicinal chemistry efforts. CADD workflows limit the need for synthesis by screening candidate molecules using a variety of approaches in order to prioritise the most promising compounds \cite{CADD_Book}. In the context of protein-ligand binding, Free-Energy Perturbation (FEP) methods based on molecular mechanics force fields \cite{Jespers2021, FEP_review} are most commonly used to assess relative free energies of binding. Yet, as shown by SAMPL blind predictive modelling challenges \cite{Rizzi2018Sample6,Amezcua2021Sample7}, state-of-the-art force field methods often yield binding free energies only moderately correlated to experimental data, depending on the chemical space, e.g. $R^2$ ranging from 0.1 to 0.9 for the OA/TEMOA dataset vs. $R^2 < 0.4$ across the board for the CB8 benchmark \cite{Rizzi2018Sample6}. Moreover, errors were found to be non-systematic and frequently exceeded 2 kcal/mol (ca. 3 mHa), thus potentially precluding meaningful ranking of molecules for the purposes of screening \cite{Rizzi2018Sample6,Amezcua2021Sample7}.  Development of low-scaling quantum-mechanical methods capable of simulating proteins is ongoing, in DFT \cite{Gundelach2021} and CC \cite{Beck2020} frameworks, but clearly, there is a pressing need for accurate methods for large-scale simulations of binding energies, not only with accurate ranking of molecules in mind, but also for the purpose of benchmarking approximate methods. 

Recently, Malone et al. presented an attempt to utilise quantum computing to compute protein-ligand binding energies, via a 1st order Symmetry Adapted Perturbation Theory (SAPT) procedure \cite{malone2021sapt}. Variational Quantum Eigensolver (VQE) quantum simulations of 6 small molecules and a subsection of the KDM5A protein were performed on a state-vector emulator to estimate their 1- and 2-electron reduced density matrices, which were then employed in a classical SAPT$^{(1)}$ calculation. However computed interaction energies did not reproduce ligand rankings yielded by more accurate \nth{2} order SAPT calculations, due to the missing induction and dispersion components in the \nth{1} order approximation \cite{malone2021sapt}.  

In this paper we compute binding energies for a series of 12 BACE1 inhibitors complexed in the active site of the BACE1 enzyme \cite{hilpert2013beta} using quantum hardware in an exhibition of what is possible with machines on the market today. The BACE1 enzyme is part of the amyloidogenic pathway and therefore associated with the pathogenesis of Alzheimer's disease \cite{Rombouts2020}. The BACE1 enzyme itself is a much-studied  system \cite{ghosh2014bace1}, and the motivations of the system selection are fourfold: availability of experimental activity data and high-resolution 3D structures of protein-ligand complexes for a series of inhibitors, limited induced fit and inhibitor modifications across the series being local to the central recognition motif \cite{hilpert2013beta}. Consequently, this system provides an ideal case for exploratory studies of protein-ligand interactions and algorithmic prototyping of fragmentation methods.

\section{Methods} \label{section:methods}

\subsection{Exploratory Model of BACE1 - ligand complexes}\label{section:model}

State-of-the-art in-silico prediction of free energies of binding usually involves a Free Energy Perturbation (FEP) protocol based on long classical molecular dynamics simulations\cite{Jespers2021}, which would be intractable if performed on a quantum-chemical level. A common approximation employed in quantum-mechanical studies of protein-ligand binding is to compute an average over single-point free energies of binding, calculated for selected configurations where the entropy and thermal corrections are obtained from a lower-level model \cite{Gundelach2021}. 

In the present study, the (electronic) binding energy, $E_{\text{bind}}$, is chosen as a simple metric by which to measure the strength of the interaction between the protein and the ligand, and is calculated as
\begin{equation}
\label{eq:metric}
    E_{\text{bind}} = E_{\text{complex (aq)}} - E_{\text{ligand (aq)}} - E_{\text{protein (aq)}},
\end{equation}
where the subscript (aq) indicates molecules in aqueous environment. We treat the protein as a set of distributed point charges and by invoking a ``frozen protein'' approximation, cancel protein-protein terms from $E_{\text{complex}}$ and simplify the above equation to a single energy difference,

\begin{equation}
\label{eq:bind}
E_{\text{bind}} = E_{\text{ligand-in-protein (aq)}} - E_{\text{ligand-in-solvent (aq)}}.
\end{equation}
The ligand is treated with a variety of electronic structure methods in the coming sections. For the treatment of solvent effects in the $E_{\text{ligand-in-protein}}$ quantity, an explicit-water solvation shell is added and the TIP3P water charges appended to the protein charges experienced by the ligand \cite{jorgensen1983comparison}. The solvation shell in $E_{\text{ligand-in-protein}}$ is not equilibrated. Solvent effects are accounted for in $E_{\text{ligand-in-solvent}}$ using the dd-COSMO implicit solvent model available in the PySCF package \cite{pyscf, lipparini2013fast, lipparini2014quantum}. The assumption is also made that the crystal structures of the ligands in the active site of each protein are sufficiently close to the optimal geometries that they need not be optimized. An optimization of the ligand in the active site under the approximations made is not possible without the parameterisation of Lennard-Jones potentials, which is beyond the scope of this publication, as the lack of Pauli-repulsion would result in the ligand being drawn unphysically close to the point charges representing the active site. In contrast, the geometries of the ligands used in the evaluation of $E_{\text{ligand-in-solvent}}$ are optimised in the presence of the implicit solvent at Hartree-Fock level using the STO--3G basis together with the geometry optimiser, such that some amount of the geometric distortion needed for the ligand to occupy the active site is accounted for in $E_{\text{bind}}$ \cite{wang2016geometry}.
The full series of oxazine ligands considered in this work is shown in figure \ref{fig:oxazines}. X-ray crystallographic structures of protein-ligand complexes are taken from prior work \cite{hilpert2013beta} and processed in MOE (Chemical Computing Group) \cite{MOE}. The 3D Protonate routine \cite{Labute2008, Labute2008jcc} was applied with standard settings. An additional 10\AA{} shell of water molecules was added and the negative charge neutralised by adding sodium cations. Protein structures were aligned with respect to the 1b protein-ligand complex. Fixed-point charges were placed on the protein atoms using the AMBER10:EHT method, as implemented in MOE. 
\subsection{Ranking metric}
The binding energy expression for ligand $i$ (eq. \ref{eq:bind}) depends explicitly on the geometries as
\[
E_{b}^{(i)}(\bar{P}_{i},W_{i})=E(\bar{P}_{i},W_{i},\bar{L}_{i})-E(\bar{W},\tilde{L}_{i})
\]
where $\bar{P}_{i}$ and $\bar{L}_{i}$ are the X-ray protein and
ligand structures for the ligand $i$, $W_{i}$ are explicitly added
water molecules, $\bar{W}$ denotes the implicit dd-COSMO water environment,
and $\tilde{L}_{i}$ is the optimized ligand geometry with dd-COSMO implicit water.
Ranking of the ligands depends on the accuracy and precision of the
metric 
\begin{align}
\begin{split}
m_{ij} &= E_{b}^{(i)}(\bar{P}_{i},W_{i})-E_{b}^{(j)}(\bar{P}_{j},W_{j})\\
&=E(\bar{W},\tilde{L}_{i})-E(\bar{W},\tilde{L}_{j}) \\ 
&\quad +E(\bar{P}_{i},W_{i},\bar{L}_{i})-E(\bar{P}_{j},W_{j},\bar{L}_{j}),
\end{split}
\end{align} 
where i and j index the system under consideration.
A more precise metric could be obtained by sampling multiple geometries.
To keep the number of quantum computations at a feasible level, but reduce uncertainties due to the explicit water we also used a modified metric with a single reference
protein-water complex of the ligand $r$ 
\begin{align}
\begin{split}
m_{ij}^{(r)}&=E_{b}^{(i)}(\bar{P}_{r},W_{r})-E_{b}^{(j)}(\bar{P}_{r},W_{r}) \\
 &=E(\bar{W},\tilde{L}_{i})-E(\bar{W},\tilde{L}_{j})\\
 &\quad +E(\bar{P}_{r},W_{r},\bar{L}_{i})-E(\bar{P}_{r},W_{r},\bar{L}_{j})
\end{split}
\end{align}
 and for the reference the protein-water complex of ligand 1b is chosen as it represents the unsubstituted prototype of the series. In other words, the energy of each protein-ligand complex is calculated using fixed point charges from the 1b-protein complex. This replacement is made possible by alignment of the protein residues, minimal induced fit and almost perfect alignment of the ligand molecules: all oxazines in the series can be seen super-imposed in the 1b protein cavity in Figure \ref{fig:ligands_in_1b_cavity_new}. We note that almost any other fixed protein configuration could have been chosen instead of that of the 1b complex. If, on the contrary, the metric uses ligand-specific protein fixed charges, the correlation to experimental data is relatively poor. A full sweep of the 12 protein structures performed at Hartree-Fock level using the STO--3G basis is  shown on Figure \ref{fig:spearmanrankings} in the Appendix. 
 
\begin{Figure}
    \centering
    \includegraphics[width=9cm]{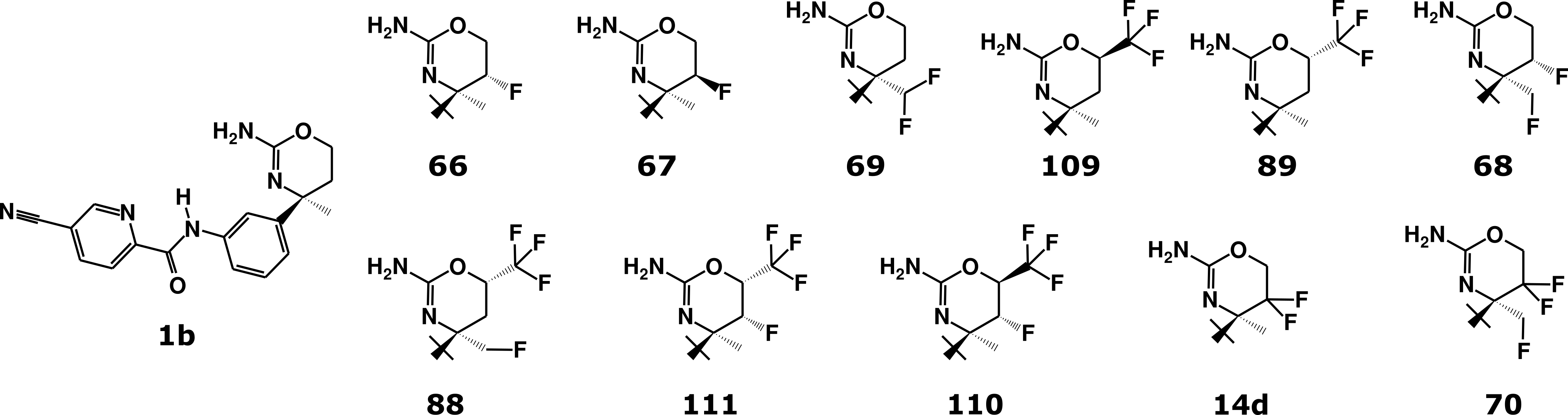}

    \captionof{figure}{Series of oxazine BACE1 inhibitors investigated by Hilpert et al. \cite{hilpert2013beta}}
    \label{fig:oxazines}
\end{Figure}

\begin{Figure}
    \centering
    \includegraphics[width=9cm]{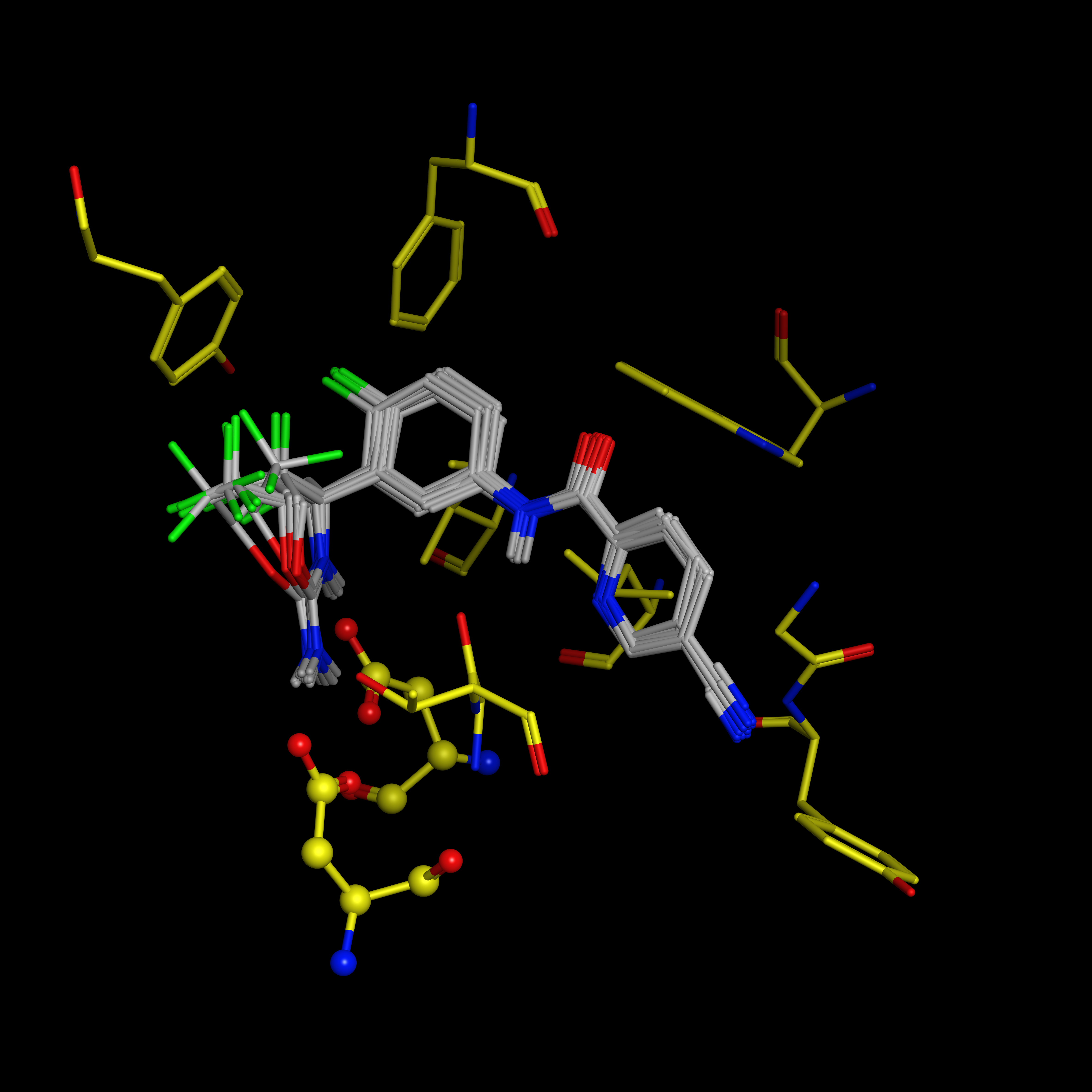}
    \captionof{figure}{Experimental ligand geometries in the 1b cavity. }
    \label{fig:ligands_in_1b_cavity_new}
\end{Figure}

For the purposes of exposition we make the assumption that there exists a linear correlation \cite{FEP_review} between the binding energy and the experimentally determined pIC$_{50}$ data \cite{hilpert2013beta} shown in table \ref{table:pIC50}.

\begin{table}[H]
\centering
\begin{tabular}{c|r}
Ligand & pIC$_{50}$ \\ \hline
89  & 7.92 \\
111 & 7.89 \\
88  & 7.72 \\
14d & 7.64 \\
66  & 7.55 \\
110 & 7.40 \\
70  & 7.31 \\
1b  & 7.29 \\
69  & 7.27 \\
68  & 7.11 \\
67  & 6.36 \\
109 & 6.11
\end{tabular}
\caption{Experimental pIC$_{50}$ data for the investigated BACE1 inhibitors shown in Figure \ref{fig:oxazines} \cite{hilpert2013beta}.}\label{table:pIC50}
\end{table}


\subsection{Density Matrix Embedding Theory}
To facilitate the simulation of such large systems on NISQ hardware, we reduce the quantum resources needed to evaluate the energy using density matrix embedding theory (DMET) \cite{knizia2013density,wouters2016practical} implemented using VQE (see next section) as a fragment solver \cite{yamazaki2018practical}. In DMET the atomic system is partitioned into fragments and the total energy of the system is the sum of the energies of fragments embedded in the environment comprising the rest of the system, represented by a quantum bath. The latter reproduces the entanglement of the embedded fragment with the environment through an interaction between $N_{\text{orb}}^A$ fragment orbitals and $N_{\text{orb}}^B \leq N^A_{\text{orb}}$ bath orbitals \cite{knizia2013density}.

The starting point of the DMET calculation is the Hartree--Fock solution of the entire molecule. In order to spatially separate the fragment from its environment we apply the Löwdin \cite{LOWDIN1970185} method to orthogonalise the atomic orbitals and use it as a localized basis for the DMET procedure. 

The Hartree--Fock (HF) Reduced Density Matrix (RDM) in the localized basis is 
\begin{equation}
    \Gamma_{pq} = \sum_{i} \bar{C}_{pi}\bar{C}^\dagger_{iq},
\end{equation}
where the quantities $\bar{C}_{pi}$ are the coefficients of the canonical molecular orbitals. The indices $p$ and $q$ refer to functions in the localised basis, and $i$ indexes the molecular orbitals. Once the localized basis orbitals of a fragment have been identified by a fragmentation strategy, the fragment bath orbitals are computed from $\Gamma_{pq}$ as described by Wouters et al. \cite{wouters2016practical}. The fragment orbitals and the bath orbitals, together referred to as embedding orbitals, are then used to construct the fragment Hamiltonian operator 
\begin{multline}
    \hat{H}_{frag}(\mu) = - \mu \sum_{p \in A} a^\dagger_{p} \hat{a}_p + \sum_{pqrs}^{N^A_{\text{orb}}} (pq|rs) \hat{a}^\dagger_p \hat{a}_r^\dagger \hat{a}_{s} \hat{a}_q \\ +\sum_{pq}^{N_{\text{orb}}^A + N_{\text{orb}}^B} ( h_{pq} + \sum^{N_\text{orb}}_{rs}( (pq|rs) - (ps|rq) ) \Gamma^{\text{env}}_{rs}) \hat{a}_p^\dagger \hat{a}_q \\
\end{multline}
where $N_{\text{orb}}^A$ is the number of orbitals on the fragment, $N_{\text{orb}}^B$ is the number of bath orbitals, $N_{\text{orb}}$ is the number of orbitals on the whole molecule, $\hat{a}^\dagger$ and $\hat{a}$ are creation and annihilation operators, respectively, $\Gamma^{\text{env}}_{rs}$ is the RDM constructed from the fully occupied orbitals in the environment, and $\mu$ is the chemical potential. The quantities $h_{pq}$ and $(pq|rs)$ are the one and two-electron integrals in chemists' notation, respectively. The fragment energy operator is
\begin{multline}
    \hat{E}_{frag}(\mu) = \sum_{p \in A} \bigg( \\ \sum_{q}^{N^A_{\text{orb}} + N^B_{\text{orb}}} \bigg(h_{pq} + \frac{\sum_{N_{\text{orb}}} [(pq|rs) - (ps|rq)] \Gamma^{\text{env}}_{rs}}{2}\bigg) \hat{a}^{\dagger}_p \hat{a}_q  \\ 
    + \frac12 \sum_{qrs}^{N^A_{\text{orb}} + N^B_{\text{orb}}} (pq|rs) \hat{a}^\dagger_p \hat{a}^\dagger_q \hat{a}_r \hat{a}_s \bigg)
\end{multline}
This procedure is performed for each fragment set by the fragmentation strategy and by using an appropriate wavefunction method the ground state $\Psi_{frag}(\mu)$ of $\hat{H}_{frag}(\mu)$ Hamiltonian is computed. Given $\Psi_{frag}(\mu)$,
\begin{equation}
    E_{frag}(\mu) = \sum_{frag} \langle\Psi_{frag}(\mu) \hat{E}_{frag}(\mu) \Psi_{frag}(\mu)\rangle
\end{equation}
is computed for each fragment and for a particular global chemical potential, $\mu$.  The value of $\mu$ is determined by the constraint 
\begin{equation}
    \sum_{frag} \langle\Psi_{frag}(\mu) \hat{N}_{frag} \Psi_{frag}(\mu)\rangle = N_{tot}
\end{equation}
where $\hat{N}_{frag}$ is the particle number operator of the fragment $frag$ and $\hat{N}_{tot}$ is the total number of electrons in the molecule. In practice, this is achieved by an iterative solver.
Finally, the DMET total energy is computed as \begin{equation}
    E = E_{nuc} + \sum_{frag} E_{frag}(\mu),
\end{equation}
where $E_{nuc}$ is the nuclear-nuclear repulsion energy in the total molecule.

\subsection{The Variational Quantum Eigensolver}
The variational quantum eigensolver (VQE) \cite{peruzzo2014variational} is a hybrid quantum/classical algorithm for the evaluation of molecular electronic ground or excited states \cite{higgott2019variational}. We use the VQE as the quantum computational solver for the DMET fragment Hamiltonian.  
 
A series of 4-qubit experiments are performed with the YXXX ansatz and a reduced active space of two electrons in four spin-orbitals. We compare the results in section \ref{section:results} with their noiseless analogues: the same calculations performed with a state vector simulator to enable comparison with ``perfect qubits''. The general VQE schema is shown in figure \ref{fig:vqe}.
\begin{Figure}
    \centering
    \includegraphics[width=9cm]{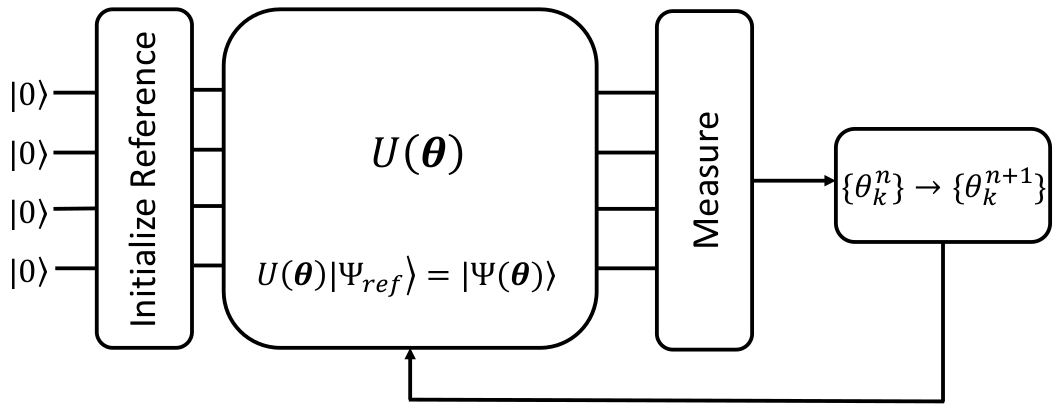}
    \captionof{figure}{VQE loop. The set of parameters $\{\theta\}$ is updated iteratively by a classical optimiser, until the measured objective function is minimized.}
    \label{fig:vqe}
\end{Figure}
The imaginary time evolution algorithm (ITE) was considered as an alternative to VQE for this work \cite{Motta2019_ITE}. However, prohibitively large numbers of shots needed to reach the desired accuracy and difficulty converging to energetic minima suggested VQE, in this instance, represented a more robust approach.

\subsection{DMET fragmentation}
The chosen fragmentation pattern is based on known interactions between the headgroup and the binding pocket \cite{hilpert2013beta}. The primary inter-atomic interactions are shown in Figure \ref{fig:bindingsite}.  Each oxazine is subjected to the same fragmentation pattern, the tail group consisting of two aromatic rings and their substituents is treated as one large fragment and is denoted in this communication by the label [TG]. The headgroup, which is host to all substitutions throughout the ligand series is split into two fragments, one which contains the atoms responsible for the primary interactions with the catalytic aspartates, the other is made up of the remainder of the headgroup including any substitutions. These are referred to as [NH$_2$CNH$^+$] and [R-HG], respectively. The [NH$_2$CNH$^+$] fragment is constant throughout the series of inhibitors being considered, but its energy is not, due to dependence on varying bath orbitals. Moreover, this fragment is the only one treated with the VQE algorithm and the YXXX ansatz. The [R-HG] and [TG] fragments are treated at Hartree-Fock level. This choice of fragmentation ensures correlation energies and, crucially, binding energies remain comparable throughout the study.
The fragmentation scheme is shown pictorially in Figure \ref{fig:fragmentation}.

\begin{Figure}
    \centering
    \includegraphics[width=5cm]{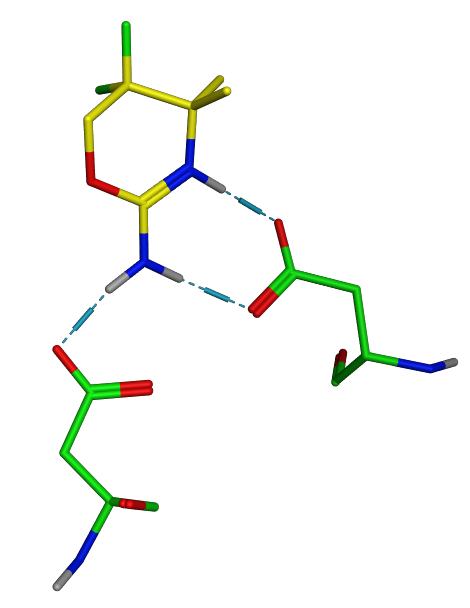}
    \captionof{figure}{The primary interactions between the ligand and ASP residues in the active site are indicated by the blue dashed markers. The tail group is omitted for image clarity.}
    \label{fig:bindingsite}
\end{Figure}

\begin{Figure}
    \centering
    \includegraphics[width=6cm]{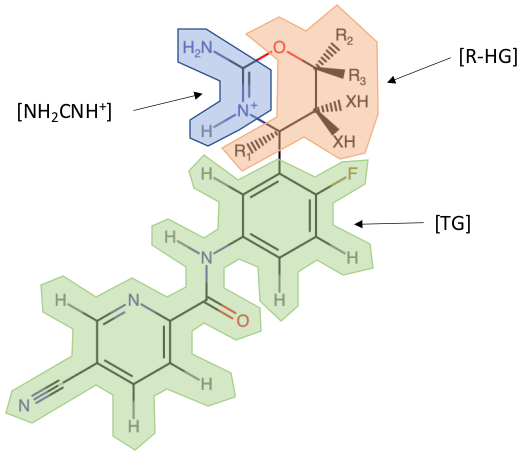}
    \captionof{figure}{The chosen fragmentation pattern for the oxazine series.}
    \label{fig:fragmentation}
\end{Figure}

\subsection{Technical details of classical calculations and quantum simulations}
All DMET calculations were performed using the EUMEN quantum chemistry software, utilizing the retargetable quantum compiler $\ensuremath{\mathsf{t}|\mathsf{ket}\rangle}$ (via its Python interface, pytket) \cite{sivarajah2020t} for quantum circuit compilation and optimization for the target platforms and PySCF \cite{pyscf} for classical Hartree--Fock and CCSD calculations as well as for evaluation of molecular orbital integrals. Qulacs \cite{suzuki2021qulacs} and Qiskit's AerSimulator \cite{Qiskit} emulators were used for the state-vector and shot-based simulations, respectively. Hardware experiments were performed on a superconducting transmon quantum computer, \textit{ibmq\_casablanca}, which is one of the IBM Quantum Falcon processors \cite{ibmq} and on a trapped-ion quantum computer, H1-S2, manufactured by Honeywell Quantum Solutions \cite{honeywell1024}. For all calculations performed on quantum hardware we use the minimal STO--3G basis set \cite{hehre1969a} and limit the active space to two electrons in four spin-orbitals. A limited depth, custom Ansatz we shall refer to hereafter as ``the YXXX Ansatz'' \cite{lee2018generalized} is used in all quantum computing simulations. The YXXX Ansatz contains only one two-electron excitation and provides a single-parameter circuit of lower depth than the analogous UCCD circuit \cite{yordanov2020iterative}. The Jordan-Wigner transformation was used to encode the molecular Hamiltonian expressed in fermionic operators into the corresponding qubit operators \cite{bauer2020quantum}.
For each energy calculation on the transmon machine, we first perform a variational calculation on the device to attain the optimised parameter for the YXXX ansatz, using 6000 shots in each iteration. The final energy value is evaluated with 60,000 shots (in 10 blocks of 6000) to reduce statistical error in the energy to that seen in the shot-based noiseless calculations. On the trapped-ion device, due to its slower speed and high qubit fidelity, we used statevector-optimised YXXX ansatz parameter and evaluated the VQE energy with 8000 shots. 
The error bars shown for the hardware and shot-based results are one standard deviation of the energy evaluated with the post-PMSV shot counts in 10 equally sized blocks with dimension N$_{s}/10$, where N$_s$ is the total number of shots surviving the symmetry verification routine.

\subsection{Error mitigation}
In this study, we use two error mitigation techniques. The first is the SPAM correction \cite{jackson2015detecting, SPAM_PhysRevA.92.042312, SPAM_Bravyi_2021} for noisy measurement and state preparation, implemented in $\ensuremath{\mathsf{t}|\mathsf{ket}\rangle}$ \cite{sivarajah2020t}. The SPAM error mitigation technique works by profiling a device by first evaluating calibration circuits involving all pairs of used qubits and comparing it with the results expected in the absence of noise to obtain a transfer matrix, the inverse of which is then applied to post-process every measurement. The second technique applied in this paper is Partition Measurement Symmetry Verification (PMSV), a recently reported flavour of symmetry verification \cite{yamamoto2021quantum}. Symmetries such as mirror planes and particle number symmetries can be represented by single Pauli strings \cite{yen2019exact}. The symmetries and Hamiltonian elements are partitioned into sets of commuting terms \cite{gokhale2019minimizing, cowtan2020generic}. If each member of a group commutes with a given Pauli symmetry, that symmetry can be used for verification of the measurement result of the corresponding measurement circuit without the need for extra resources, i.e. any measurement which violates the known symmetry of the system must be due to noise, and hence is discarded during post-processing. 

\section{Results} \label{section:results}
We have constructed a simplified model for calculation of binding energies of ligands with the binding site of the BACE1 protein. We first briefly review two literature reports on theoretical calculations of ligand-BACE1 binding energies. We evaluate our model using classical mean-field methods and subsequently implement it within a quantum computing framework. 

\subsection{State-of-the-art in computation of ligand-BACE1 binding energies}

Binding energies of 35 small molecules with monocyclic and bicyclic amidine or guanidine cores to a simplified model of the BACE1 protein have been calculated using B3LYP and M06-2X functionals in the 6-31+G$^{**}$ basis set and compared to experimentally determined potencies (pIC$_{50}$) \cite{Roos2014DFT_BACE1}. To compute the binding energies, the authors used single-point electronic energy calculations of solvated ligand and ligand-protein complexes, neglecting entropy effects. A fair, solvation model-dependent correlation between predicted and experimental ligand affinities was observed, with the SM8 model yielding a $R^2$ of 0.65 and Mean Absolute Deviation (MAD) of 0.4 kcal/mol (0.6 mHa). However, the predicted relative binding energies were, according to the authors, exaggerated compared to experimental binding energies, with the slope of core potency to predicted affinity of about 0.4 (instead of the expected 1.0) \cite{Roos2014DFT_BACE1}. The accuracy of this DFT-based model was similar to a state-of-the-art classical molecular dynamics-based Free Energy Perturbation (FEP) simulation of relative binding affinity of 32 BACE1 inhibitors to the full BACE1 protein ($R^2$ 0.7, MAD 0.6 kcal/mol).\cite{Ciordia2016_BACE1_FEP} These two studies demonstrate that state-of-the-art methods offer only modest accuracy and that single-point energy calculations on a simplified system can yield comparable results to FEP simulations of the whole protein-ligand complex.

\subsection{Evaluation of the Model with Classical Methods}

To better understand the limitations of the present model, we compute the binding energy of all ligands in the series in the binding pocket of 1b as described in the Methods section,  using  a  density  functional of comparable or better quality \cite{GMTKN}, $\omega$B97X \cite{wb97x},  to  the  one used in the study by Roos \textit{et al.} \cite{Roos2014DFT_BACE1} cited in the previous section and a larger basis set, def2-TZVP, as well as at the HF/STO--3G level. The results of these calculations are shown in Figure \ref{fig:Ebinds 1b} and demonstrate that application of a large basis set and a more sophisticated electronic structure method does not improve the agreement of binding energy ranking with the experimental ranking compared to HF/STO--3G. Hence, the use of a minimal basis set is not likely to be the main source of error in our model. 

\begin{Figure}
    \centering
    \includegraphics[width=9cm]{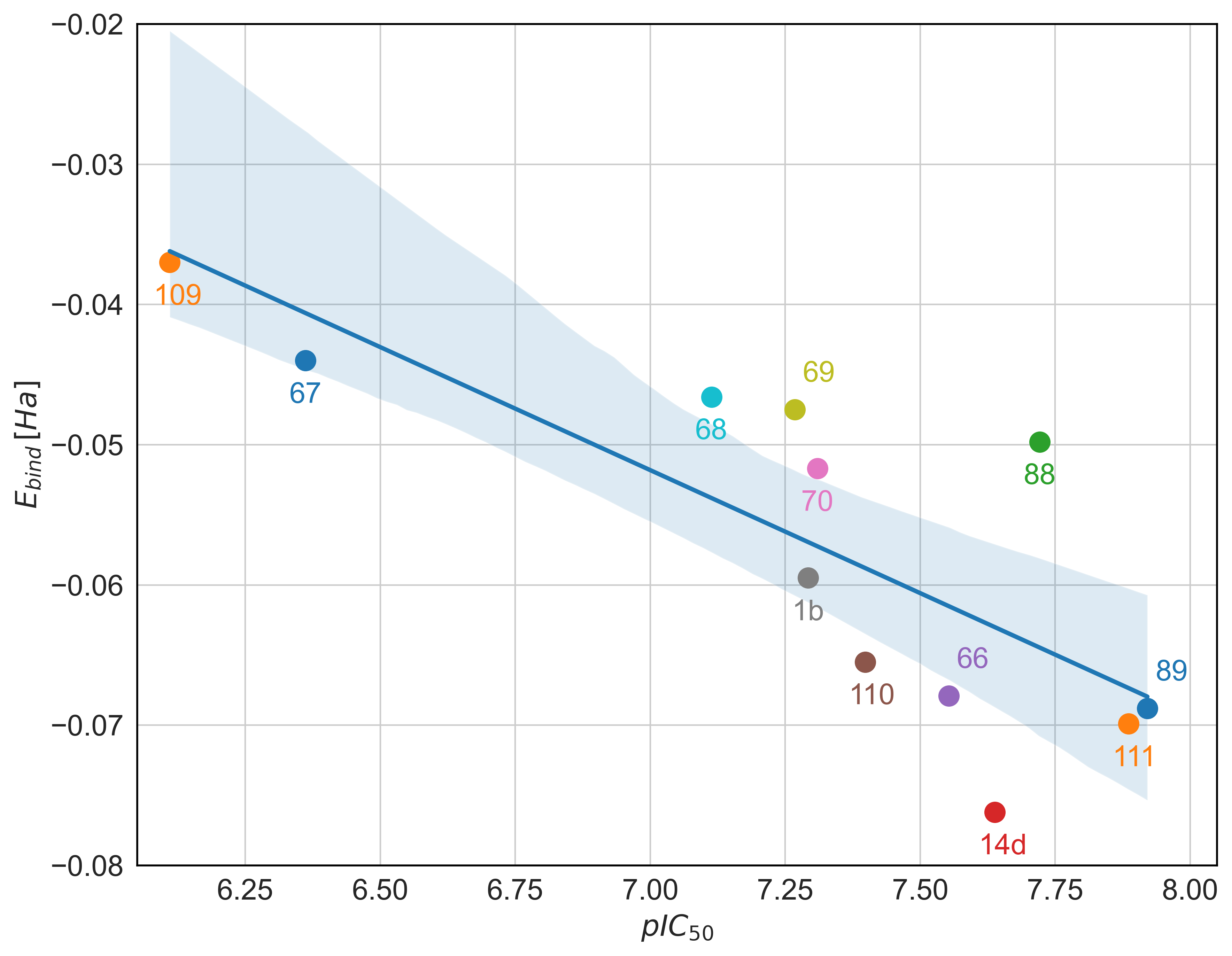}
    \includegraphics[width=9cm]{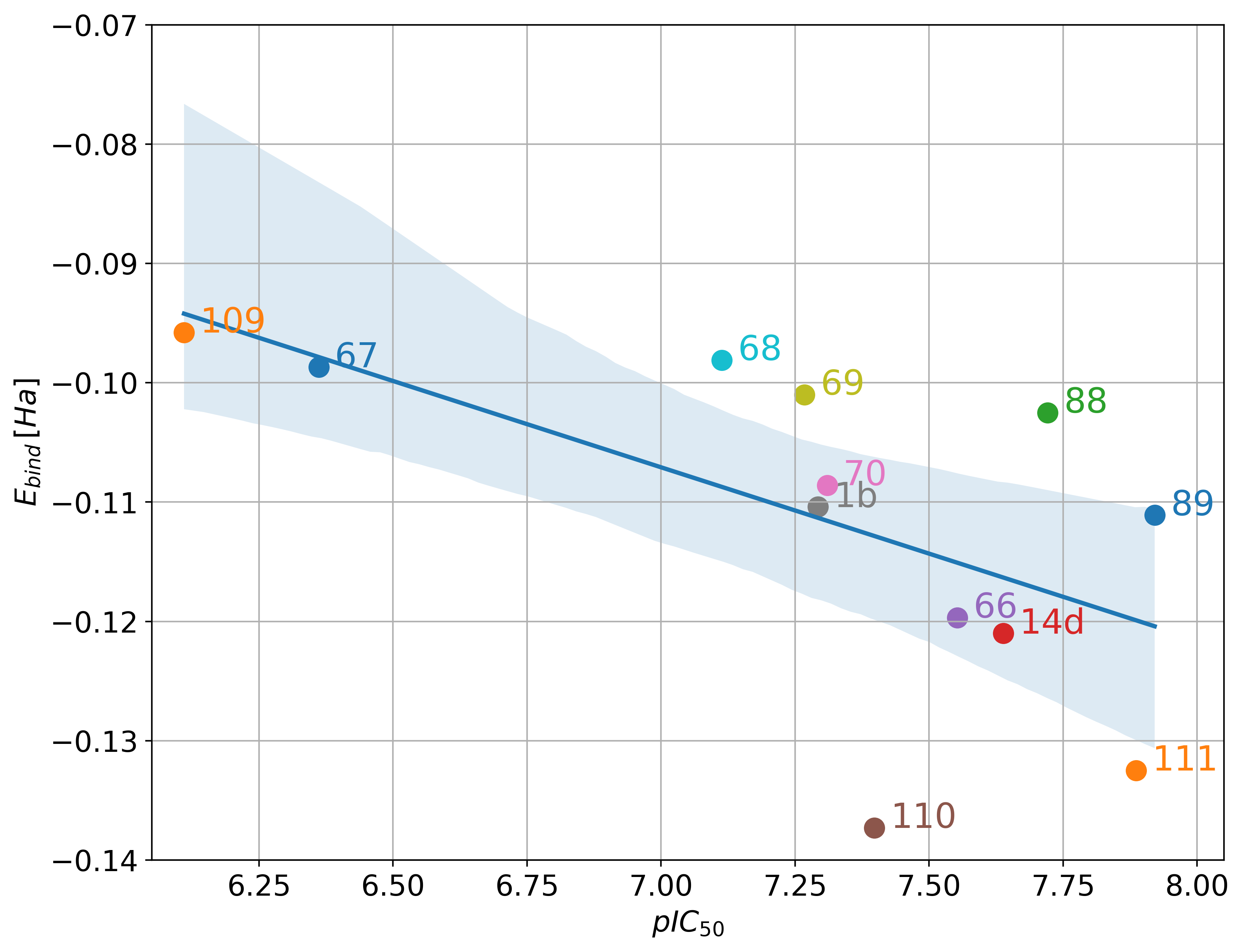}
    \captionof{figure}{RHF/STO--3G (top, $R^2 = 0.61$) and $\omega$97X/def2-tzvp (bottom, $R^2 = 0.35$) binding energies against experimental pIC$_{50}$. Each point is labelled with the corresponding ligand ID.}
    \label{fig:Ebinds 1b}
\end{Figure}

\subsection{Classical DMET Calculations and Noiseless Quantum Simulations}
Adopting the fragmentation strategy outlined in the Methods section, we construct the full molecular energy as a sum of the fragment energies for the ligand both in solvent and in the active-site of the BACE1 protein and compute the binding energy according to equation \ref{eq:bind}. To establish a baseline for DMET experiments on quantum computers, we have performed a series of calculations where the solver for the [NH$_{2}$CNH$^+$] fragment is either classical Hartree-Fock, classical CCSD (with or without an active space approximation) or state-vector VQE simulation on a quantum emulator. For the [TG] and [R-HG] fragments, the solver is always chosen to be classical Hartree-Fock.

\begin{figure*}[t]
  \begin{minipage}[t]{0.5\linewidth}
    \centering
    \includegraphics[width=9cm]{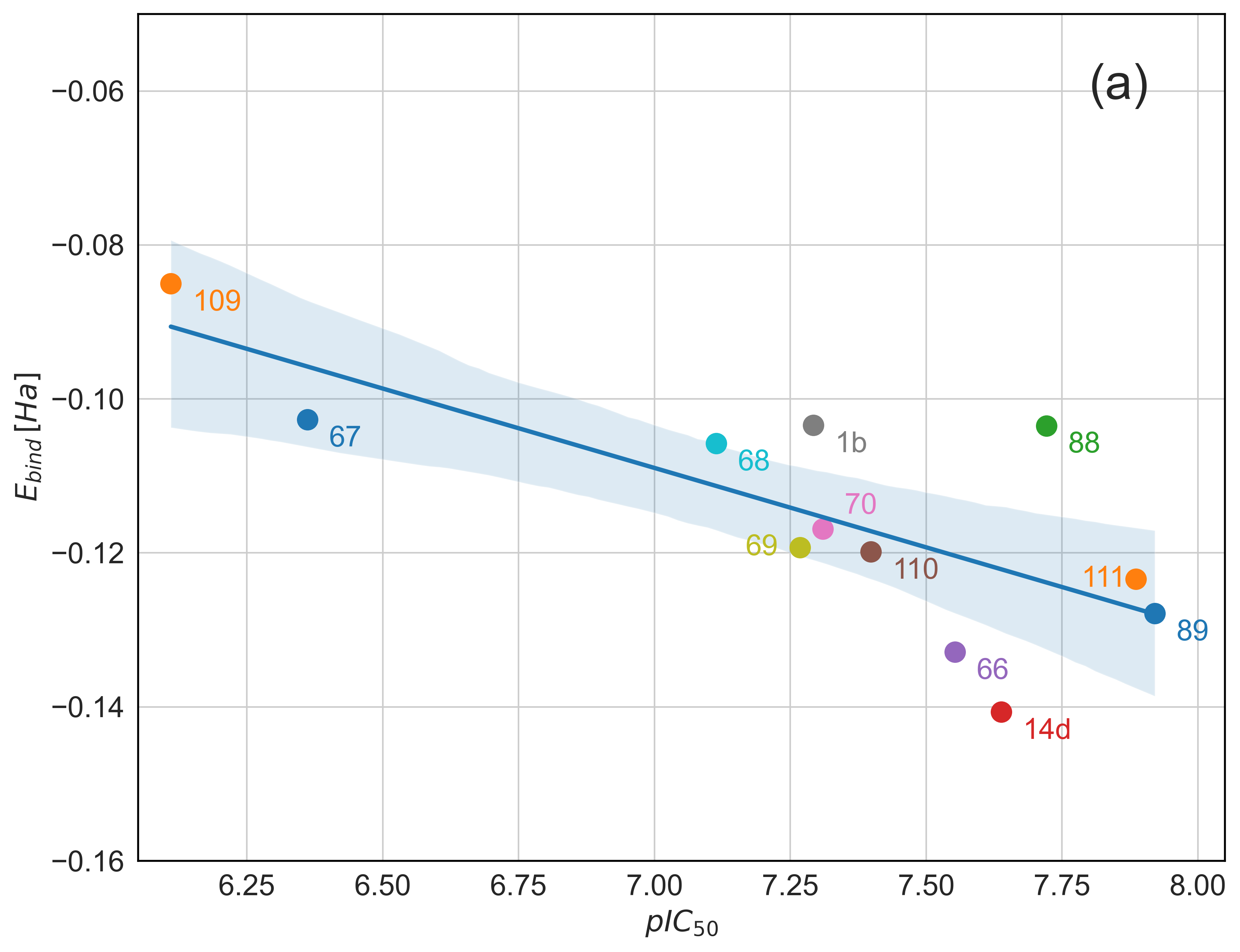} 

  \end{minipage}
  \begin{minipage}[t]{0.5\linewidth}
    \centering
    \includegraphics[width=9cm]{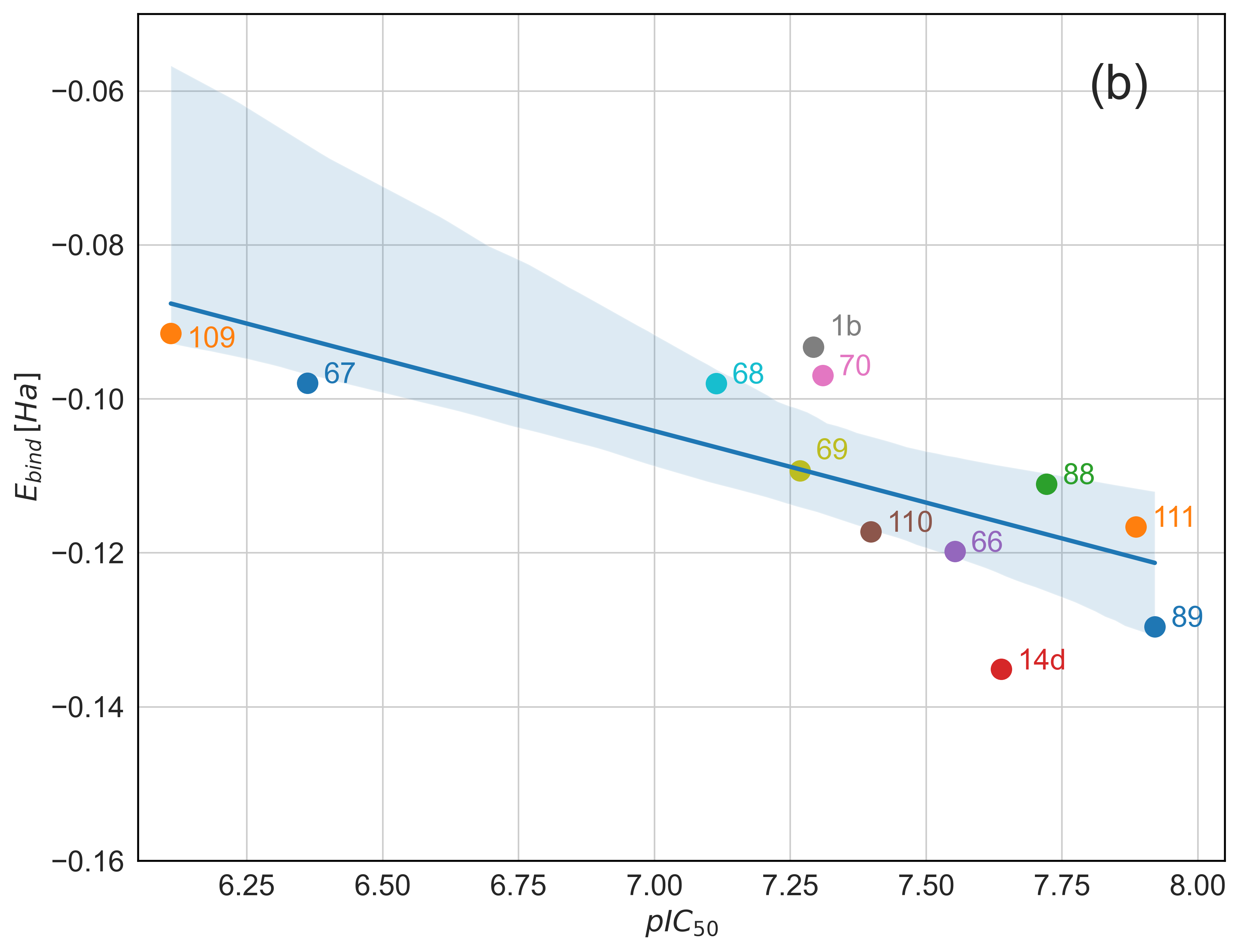} 
  \end{minipage} 
  \begin{minipage}[t]{0.5\linewidth}
    \centering
    \includegraphics[width=9cm]{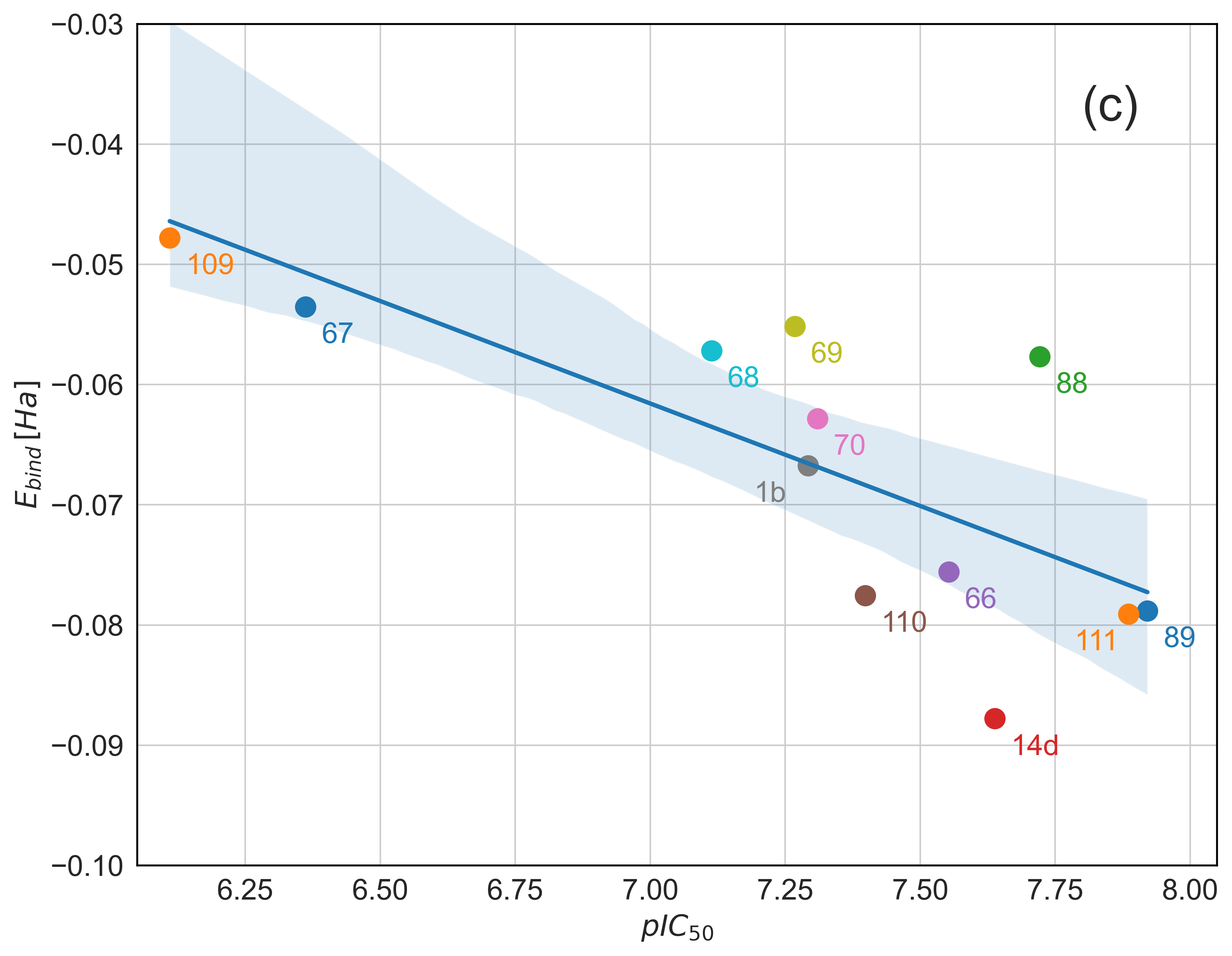} 
  \end{minipage}
  \begin{minipage}[t]{0.5\linewidth}
    \centering
    \includegraphics[width=9cm]{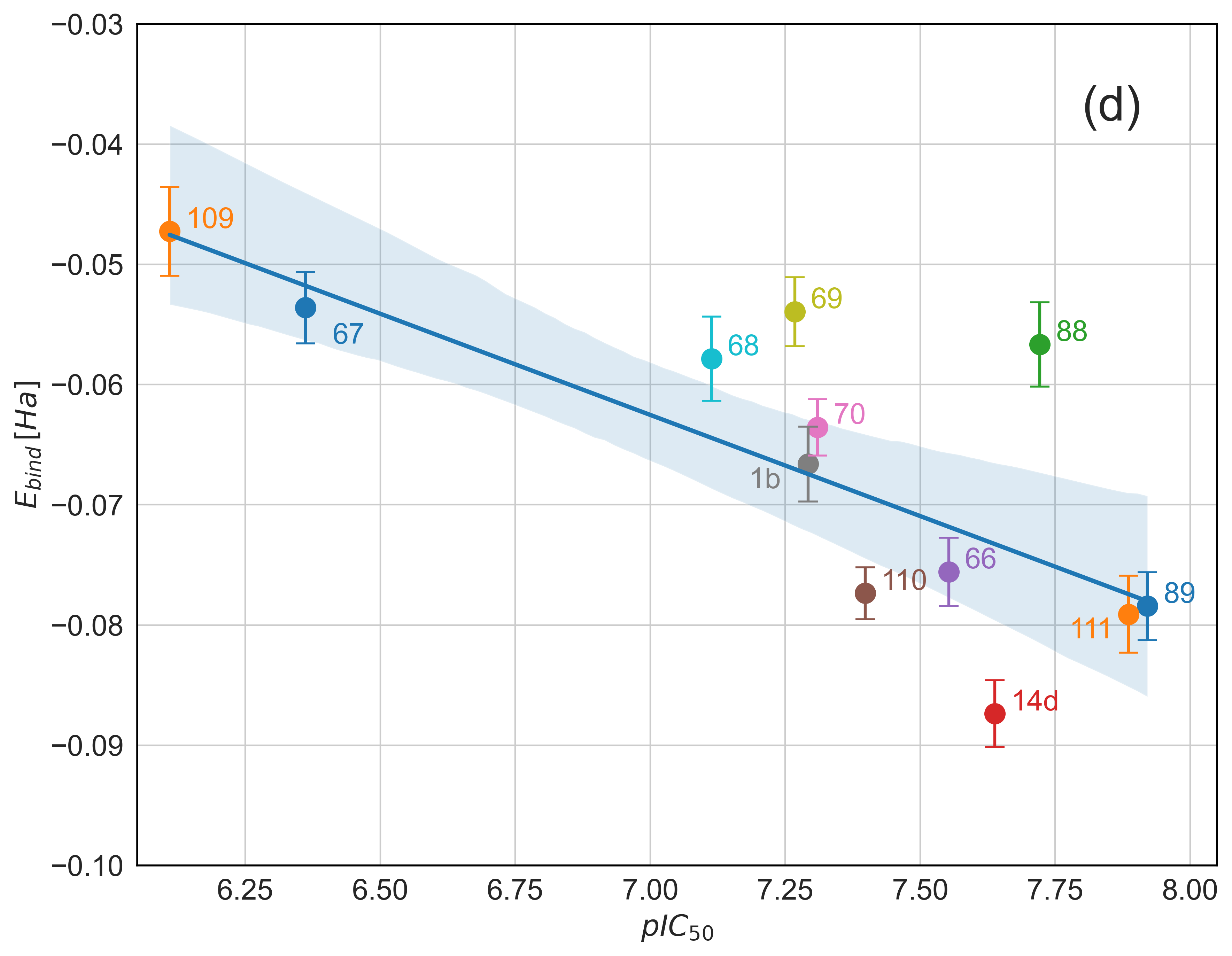} 
  \end{minipage} 
  \captionof{figure}{Binding energies obtained using (a) a classical CCSD solver without an active space, (b) a classical CCSD solver with a HOMO/LUMO active space, (c) a state vector YXXX-VQE solver with a HOMO/LUMO active space, and (d) a shot-based, noiseless YXXX-VQE solver with a HOMO/LUMO active space for the [NH$_{2}$CNH$^+$] fragment. All energies are shown in Hartrees, the light blue shaded regions represent the 95\% confidence intervals. In each graph the points are labelled with the corresponding ligand IDs. 
  }
  \label{fig:classical Ebind plots}
\end{figure*}

We first consider a CCSD classical solver for the [NH$_{2}$CNH$^+$] fragment both with and without an active space approximation. We show in figure \ref{fig:classical Ebind plots}a the binding energies obtained using CCSD with no active space restriction. Figure \ref{fig:classical Ebind plots}b shows the binding energies calculated using CCSD with a 2-electrons-in-4-spin-orbitals active space (to match the VQE simulations with the YXXX Ansatz) and demonstrates minor errors introduced by the active space approximation: the ordering of some strongly-bound ligands is reversed, but the overall trend is maintained. 

Before appealing to hardware we first evaluate the series of binding energies using a state-vector simulator, which provides a noiseless benchmark for the quantum device calculations described in the coming subsection. The results of these simulations are shown in Figure 
\ref{fig:classical Ebind plots}c.

The outcome of running the VQE calculation is an approximation to the electron correlation energy, which, by definition, must be negative. With noisy quantum devices, it is not guaranteed to be the case: one could feasibly measure an energy above the mean-field result as a consequence of  device errors. As such, to facilitate meaningful discussion in the later sections and benchmark the device errors we experience, we show in table \ref{table:correlation energies} the correlation energies obtained from the state vector simulations. It is important to note that the correlation energies do not cancel out in the difference between energies of bound and solvated ligands. On the contrary, the interaction of the [NH$_2$CNH$^+$] fragment with the charge shell representing the protein results, in all cases, in a larger correlation energy. Since electron correlation depends on the separation between electrons, it is likely that a large portion of this correlation arises through the positively charged fragment drawing in electron density from the rest of the molecule. 

\begin{center}
\begin{tabular}{c|rrr}
Ligand & $E^{\text{corr}}_{\text{ligand-in-protein (aq)}}$ & $E^{\text{corr}}_{\text{ligand (aq)}}$ \\ \hline
89 &	-0.0345 &	-0.0245 \\
111 &	-0.0354 &	-0.0262\\
88 &	-0.0324 &   -0.0246\\
14d &	-0.0348 &	-0.0232\\
66 &	-0.0294 &	-0.0218\\
110 &	-0.0386 &	-0.0266\\
70 &	-0.0345 &	-0.0234\\
1b &	-0.0273 &	-0.0201\\
69 &	-0.0282 &	-0.0205\\
68 &	-0.0326 &	-0.0220\\
67 &	-0.0313 &	-0.0218\\
109 &	-0.0358 &	-0.0252\\
\end{tabular}
\captionof{table}{Correlation energies obtained using the YXXX ansatz in state-vector calculations of $E_{\text{ligand-in-protein (aq)}}$ and $E_{\text{ligand (aq)}}$ for each oxazine.}\label{table:correlation energies}
\end{center}

Figure 
\ref{fig:classical Ebind plots}d shows the results of shot-based noiseless calculations of all 12 oxazines binding energies. These quantities provide an estimate for what proportion of the uncertainty seen in the hardware simulations arises through the limited sampling of the distribution of states. In total we perform 60,000-80,000 shots per each energy evaluation and, on average, reduce the deviation from state vector simulations to less than 0.002 Ha. 

\subsection{Simulations on Quantum Computers}
We perform all hardware experiments using IBM's \textit{ibmq\_casablanca} superconducting transmon and Honeywell Quantum Solutions' \textit{H1-S2} trapped-ion NISQ devices. We find that application of error mitigation techniques is necessary for obtaining meaningful results of the simulation. Without error mitigation, we observe large variance of results and significant deviations from expected energies, in the worst case resulting in a shift of almost $0.77$ Ha away from the state vector result, while in the best case we see a disagreement of only $0.012$ Ha. Post-processing the device outputs with PMSV method, or with both PMSV and SPAM, results in suppression of errors to $\pm$ 0.02 Ha, see Figure \ref{fig:Efrag mitigation comparison COSMO QMMM}. Comparison with Table \ref{table:correlation energies} shows that using the raw (unmitigated) quantum device shot counts to construct the total energies leads to errors similar to, or in some cases larger than, the correlation energies for the oxazines in the series. This highlights the need for an effective error mitigation approach. 

Figure \ref{fig:Efrag mitigation comparison COSMO QMMM} shows the impact of the post-processing of measurement counts using both regimes for both the fragment energy in the implicit solvent, and in the charge shell representing the protein, computed on the \textit{ibmq\_casablanca} device.  We observe that the maximum deviation of the fragment energy from the state vector result is $0.019$ Ha (an improvement over the unmitigated results by an order of magnitude), while the minimum deviation from the ideal result is just $0.0004$ Ha. In all cases we observe that the deviation from the state vector result is less than the correlation energy of the respective fragment. 

\begin{Figure}
    \centering
    \includegraphics[width=9cm]{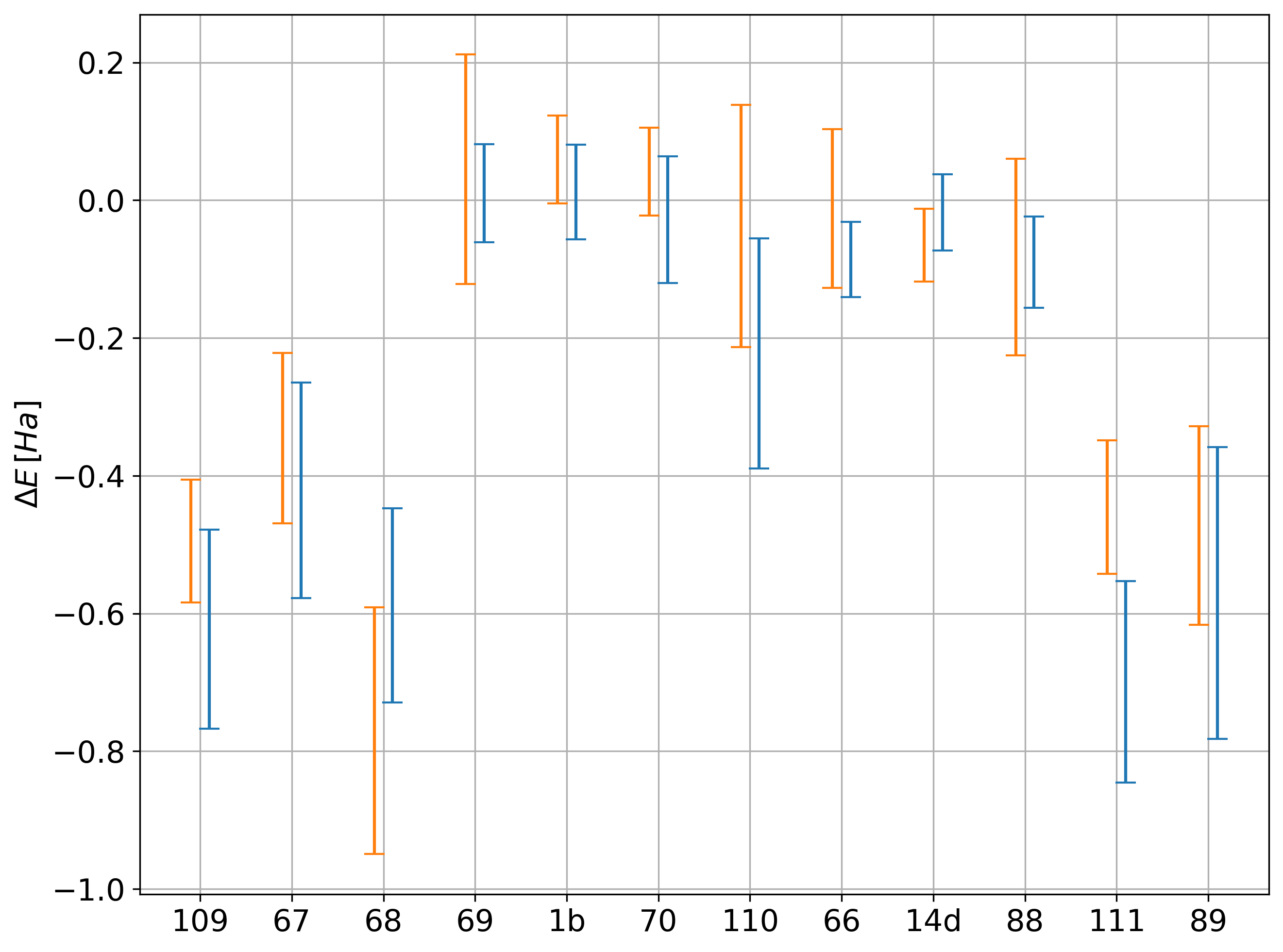}
    \includegraphics[width=9cm]{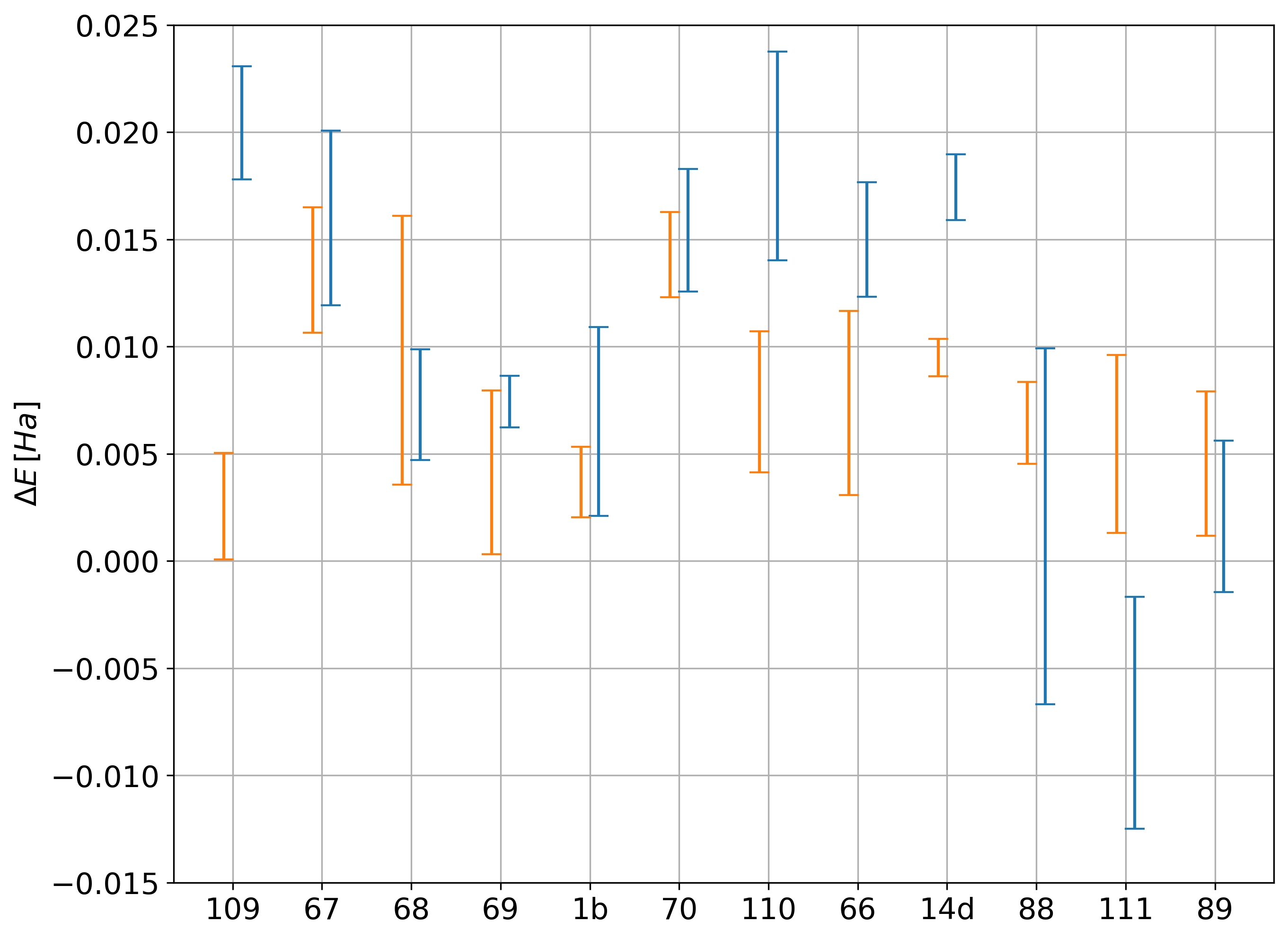}
    \includegraphics[width=9cm]{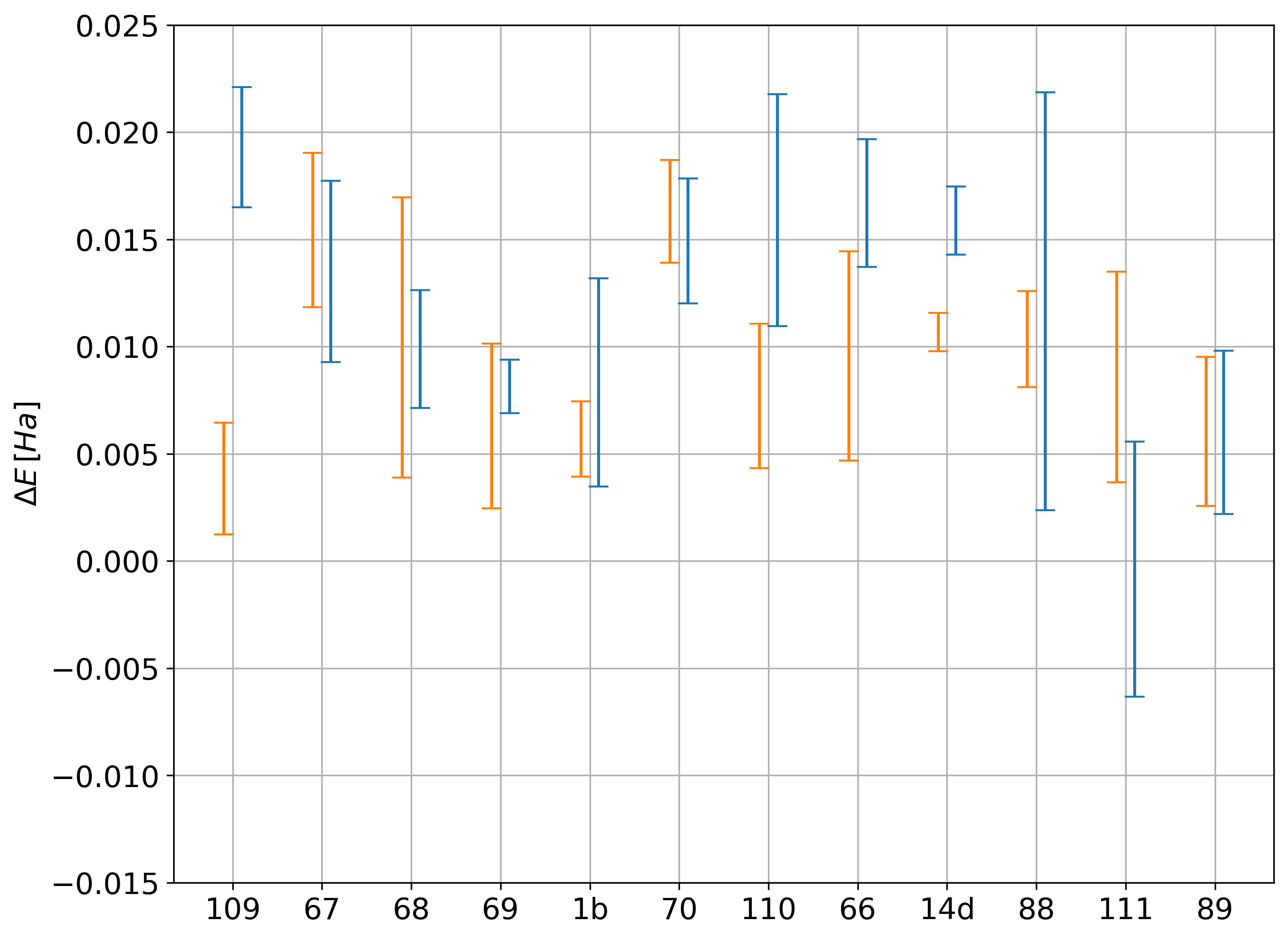}
    \captionof{figure}{Deviations of fragment energies calculated on the ibmq\_casablanca hardware (HW) from statevector (SV) simulations (in Hartree) of $E_{\text{ligand (aq)}}$ (orange line) and $E_{\text{ligand-in-protein (aq)}}$ (blue line), without error mitigation (top), after the application of PMSV in isolation (middle) and in conjunction with SPAM (bottom). The error bars represent standard deviations of data.}
    \label{fig:Efrag mitigation comparison COSMO QMMM}
\end{Figure}

These results show that successive application of the SPAM and PMSV error mitigation techniques does not yield significant improvement over using PMSV on its own. Hence, we elect to only use PMSV for further experiments on the transmon and trapped-ion devices. In Table \ref{table:comparisson_ibmq_hqs}, we compare deviations of binding energies obtained from hardware experiments on transmon (all 12 ligands) and trapped-ion devices (six selected  ligands, in order to most effectively utilise limited computational resources) from ideal state-vector results and note similar accuracy delivered by both device architectures under our experimental conditions. 
Figures \ref{fig:Ebind YXXX casablanca} and \ref{fig:Ebind YXXX HQS_H1_2} show the binding energies plotted against pIC$_{50}$ calculated on \textit{ibmq\_casablanca} and \textit{H1-S2} after PMSV is applied. 
\begin{Figure}
    \centering
    \includegraphics[width=9cm]{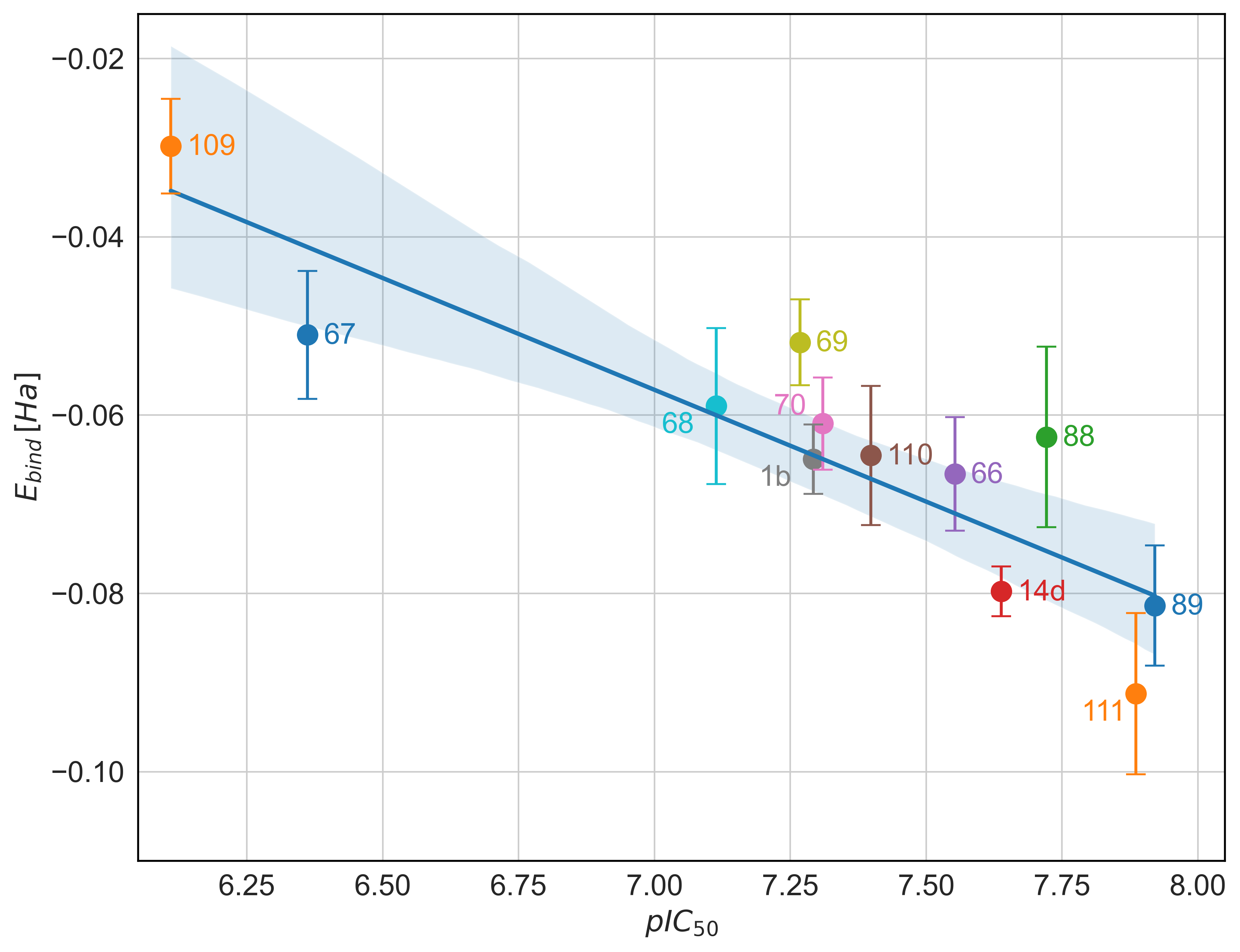}
    \captionof{figure}{Binding energies $E_{bind}$ (in Hartree) computed on ibmq\_casablanca using DMET-VQE with the YXXX ansatz. The solid line is a line of best fit ($R^2 = 0.77$), the light-blue area represents 95\% confidence interval. The error bars are standard deviations of the results. Each point is labelled with the corresponding ligand ID.}
    \label{fig:Ebind YXXX casablanca}
\end{Figure}

\begin{Figure}
    \centering
    \includegraphics[width=9cm]{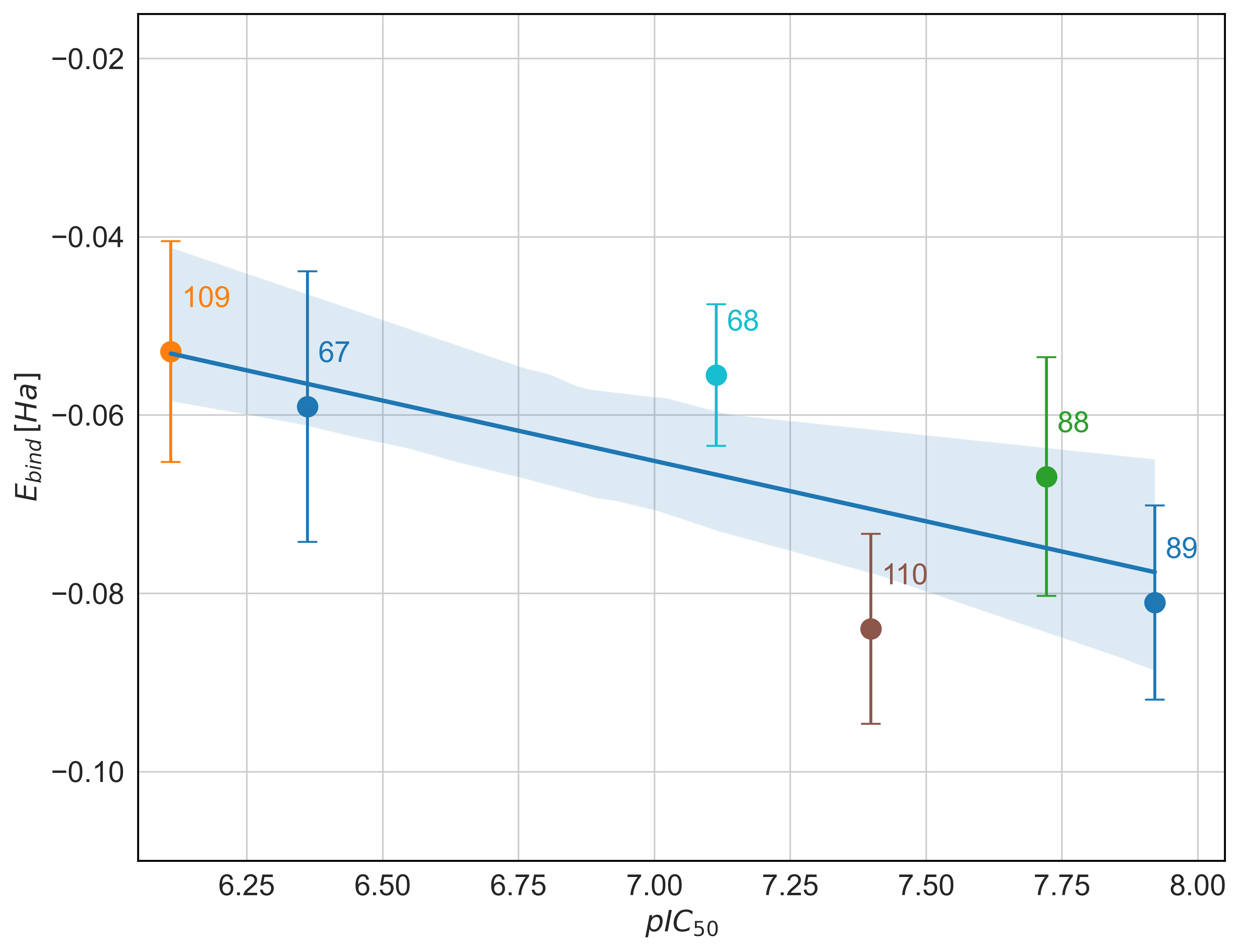}
    \captionof{figure}{Binding energies $E_{bind}$ (in Hartree) computed on H1-S2 device using DMET-VQE  with the YXXX ansatz. The solid line is a line of best fit ($R^2 = 0.56$), the light-blue area represents 95\% confidence interval. The error bars are standard deviations of the results. Each point is labelled with the corresponding ligand ID.}
    \label{fig:Ebind YXXX HQS_H1_2}
\end{Figure}

\begin{center}
\begin{tabular}{c|rrr}
Ligand &$E_{\text{bind}}^{\text{Statevector}}$ & $\Delta E_{\text{bind}}^{\text{Transmon}}$  & $\Delta E_{\text{bind}}^{\text{Trapped-Ion}}$ \\ \hline
89 & 	-0.07883 &-0.00254	& -0.00220 \\
111 &	-0.07909 &-0.01217 & - \\
88 &	-0.05767 &-0.00479 & -0.00922 \\
14d &	-0.08779 & 0.00803 & - \\
66 &	-0.07558 & 0.00899 & - \\
110 &	-0.07755 & 0.01303 & -0.00642 \\
70 &	-0.06284 & 0.00189 & - \\
1b &	-0.06673 & 0.00179 & - \\
69 &	-0.05515 & 0.00332 & - \\
68 &	-0.05718 &-0.00180 & 0.00168 \\
67 &	-0.05353 & 0.00253 & -0.00551 \\
109 &	-0.04778 & 0.01795 & -0.00510\\
\end{tabular}
\captionof{table}{Deviations (in Hartree) of binding energies computed from hardware experiments performed on transmon (IBM) and trapped-ion (Honeywell) devices from statevector simulation results.}\label{table:comparisson_ibmq_hqs}
\end{center}

\subsection{Ranking of BACE1 ligands}
One of the main objectives of calculation of protein-ligand binding energies is ranking of ligands to support the prioritisation of design ideas \cite{FEP_review}. 
Ideally, the simulated ranking should reproduce experimental trends in potencies, such as pIC$_{50}$ values. We used coefficients of determination $R^2$ of pIC$_{50}$ and computed binding energies to assess the quality of our model. For the ideal state-vector simulations, $R^2$ equals 0.55, indicating modest agreement with experimental data. Simulations obtained on \textit{ibmq\_casablanca} yield a $R^2$ of 0.77, indicating non-negligible impact of device noise on the ligand order and fortuitous error cancellation. Quantum simulations performed on the \textit{H1-S2} device on half of the ligands result in a $R^2$ of 0.56, essentially identical to that of the state-vector simulation.  

The ranking we obtained on the quantum computer is not improved with respect to the one calculated at HF/STO--3G level (see above), presumably due to the simplicity of our model of the protein-ligand interaction, which does not guarantee convergence to the experimental data. We note, however, that our results are comparable to the correlation of the data reported in the DFT study by Roos et al. \cite{Roos2014DFT_BACE1} ($R^2 = 0.65$). 

Aside from the coefficient of determination, one can assess the quality of our prediction by comparing the distributions of binding energies computed for the ``weakly bound ligands'' (67 and 109) to the ``strongly bound'' ones (the remaining molecules), see Figure \ref{fig:ligand discrimination}: while the distributions are not perfectly separated, their means are shifted by over 0.02 Ha. 

\begin{Figure}
    \centering
    \includegraphics[width=9cm]{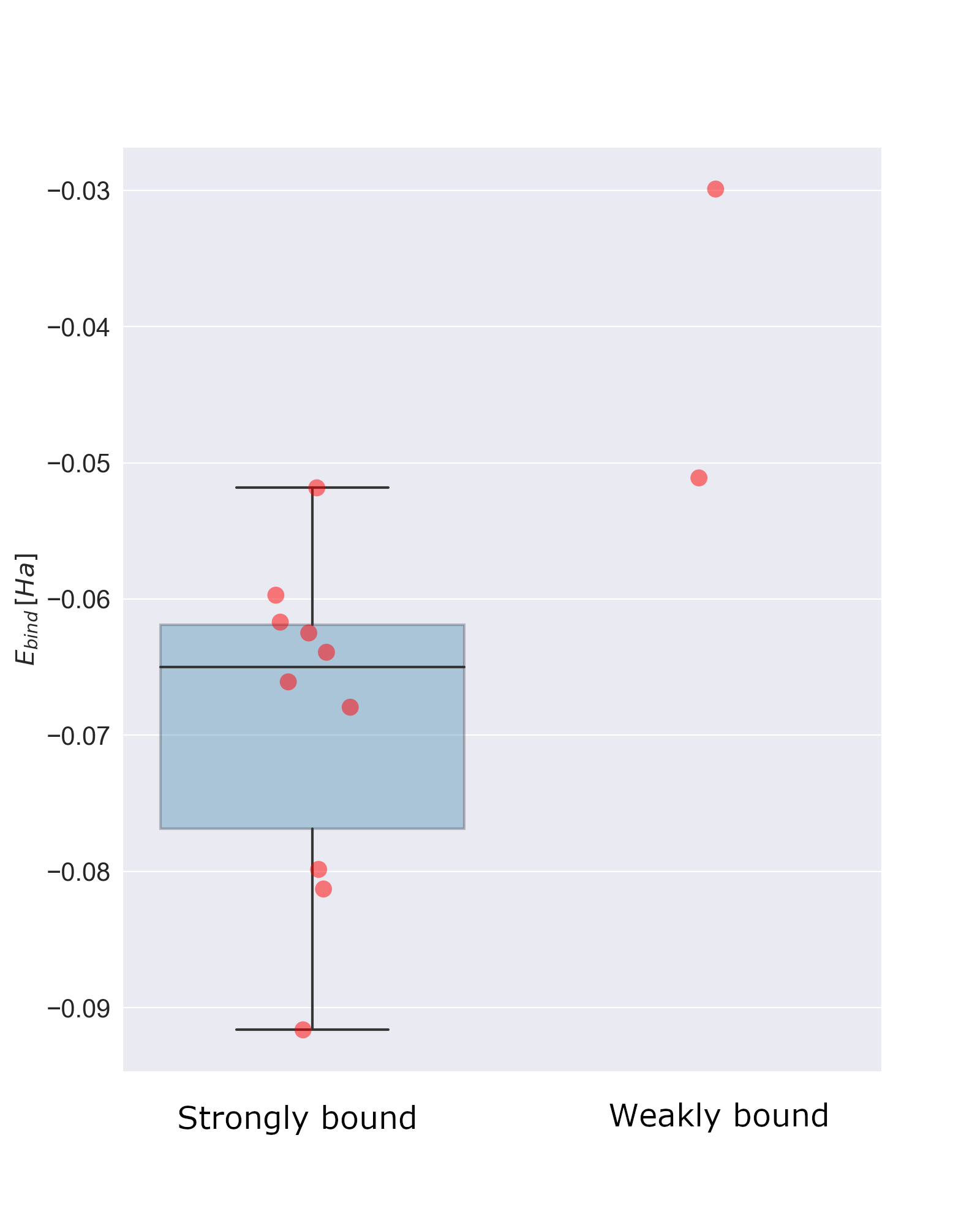}
    \captionof{figure}{Comparison of distributions of binding energies, computed on \textit{ibmq\_casablanca}, of strongly bound  and weakly bound ligands (see text).}
    \label{fig:ligand discrimination}
\end{Figure}

\section{Discussion} \label{section:discussion}
In this communication we provided a simple model and the first application of quantum simulation to pharmaceutically interesting chemical systems. We took a much-studied system for which benchmark data exists, and showed that by making simplifications and using embedding techniques, it is possible to make use of NISQ hardware to compute protein-ligand binding energies. While the ranking of ligands by the computed binding energies does not reproduce experimental ordering, it is comparable (in terms of the correlation coefficient) to results of state-of-the-art DFT and classical MD simulations on similar BACE1 inhibitors reported by others \cite{Roos2014DFT_BACE1,Ciordia2016_BACE1_FEP}. 

Our approach differs significantly from that of Malone et al. \cite{malone2021sapt}, who also used quantum computation to estimate protein-ligand binding energies. Their simulations were, however, performed on a classical quantum emulator. They claim that their method, SAPT(VQE), is inherently more precise due to reliance only on a single VQE measurement, as opposed to calculating the difference between two VQE energies, because the latter results in accumulation of errors. However, the accuracy of Malone's workflow is limited by truncation of the SAPT expansion at \nth{1} order and it is not clear how their method can be effectively extended to higher orders.

The method presented in this publication, while hostage to many approximations and not production-ready yet, has a strong connection to established embedding and QM/MM methods and can, therefore, be systematically improved. It would be possible to extend this model by taking many snapshot calculations as each protein-ligand complex is propagated classically, or to refine its QM part by including atoms of the protein's binding site in the QM region, applying a correlated method to more fragments and including van der Waals interactions between the QM fragment and the rest of the system. The [NH$_2$CNH$^+$] fragment is among the largest (in terms of number of atoms) chemical systems simulated on quantum hardware to date, the results obtained from these experiments show promisingly small amounts of noise after PMSV error mitigation is applied. It is worth emphasising that without PMSV error mitigation, deviations from ideal state-vector results are very large and far exceed the correlation energies themselves.

A major bottleneck in the application of quantum computation to the evaluation of molecular energies is found in the circuit depth required to construct the quantum state corresponding to chemical Ans{\"a}tze, which very quickly becomes intractable on NISQ machines as active spaces expand to many qubits. Relatively high gate times (in comparison to transistor times in classical computers) lead to the requirement that the state of the qubit register remains coherent over time scales not currently possible. With parallel gate execution quickly becoming a reality, advances in quantum error correction and the continued improvement of QPUs in other aspects, we expect the size and complexity of models such as the one in this publication to increase rapidly, and the number of necessary approximations to facilitate quantum simulation to decrease. Specifically, we expect that quantitatively accurate results could be obtained if (1) a high-quality (e.g. triple-$\zeta$) atomic basis set and a large (or even complete) active space are used, (2) a correlated calculation is performed on several (if not all) fragments and (3) at least the aminoacid residues closest to the ligand are included in the QM region. The first condition would warrant using up to 250 qubits for the [NH$_2$CNH$^+$] fragment, necessitating significant improvements in capabilities of quantum hardware compared to current state-of-the-art, as well as novel software techniques to reduce quantum resource requirements. Such methods may comprise e.g. a combination of a compact qubit mapping \cite{kirby2021secondquantized} and PNO approximation \cite{Kottmann_2021}, or application of a transcorrelated Hamiltonian \cite{Motta_transcorrelated_2020}. Treatment of all fragments at a correlated level would be straightforward, as long as fragment size is kept limited (i.e. comparable to [NH$_2$CNH$^+$]). The requirement of having an explicit representation of the closest protein residues can be met by, for example, following the recipe for a minimal BACE1 active site model of Roos et al. \cite{Roos2014DFT_BACE1}.   Furthermore, while our workflow currently uses VQE executed on a NISQ device, there is nothing to prevent it from using a Quantum Phase Estimation solver running on a fault-tolerant quantum computer, once the latter is available. 
Precise evaluation of interaction energies, whether between small molecules or between proteins and ligands, is one of the central problems in computational chemistry and has countless practical applications. Hence, we expect the model presented herein to provide a platform for the development of more refined models as the capabilities of quantum computing hardware improve.

\section{Conclusion} \label{section:conclusion}
This exploratory study of protein-ligand binding energies evaluated using quantum hardware, as far as we are aware, is the first of its kind and provides an indicator of how rapidly the field of quantum computation has progressed in recent years. The purpose of this publication is to lay the groundwork for future development of a productive workflow for CADD and to evaluate the performance of hardware when the energies being measured are to the order of hundreds of Hartrees. We conclude that, if appropriate error mitigation techniques are applied, and the system simplified sufficiently, NISQ hardware is capable of recovering the trend obtained using classical computers. We suggest that improvement of the workflow described herein need not be limited to progress in hardware capabilities but also  found in the refinement of the model, by expansion of the QM region to include atoms of the binding site as well as inclusion of dispersion interactions between the QM and MM regions.

\section*{Acknowledgements}
We acknowledge the use of IBM Quantum services for this work. The views expressed are those of the authors, and do not reflect the official policy or position of IBM or the IBM Quantum team.
We thank Brian Neyenhuis and all the Honeywell Quantum Solutions team for their availability and support with the H1-S2 device. The state-vector simulations in this work were performed on Microsoft Azure Virtual Machines provided by the ``Microsoft for Startups'' program. We would also like to thank Irfan Khan (CQ) for his helpful advice regarding hardware experiments. We thank Georgia Christopoulou and Andrew Tranter (CQ) for reviewing the manuscript of this work and for their valuable comments. We are grateful to Alyn Chad Edwards (CQ) for project management and providing feedback on the manuscript draft. We thank the members of the Roche Pharma Quantum Computing Task Force, specifically Mari\"elle van der Pol, Stanislaw Adaszewski and Yvonna Li for insightful discussions and Bryn Roberts for initiating this Task Force at Roche. 

\printbibliography

\section{Appendix}\label{section:appendix}
\subsection{Fixed or ligand--specific protein charges?}
\begin{figure}[H]
    \centering
    \includegraphics[width=9cm]{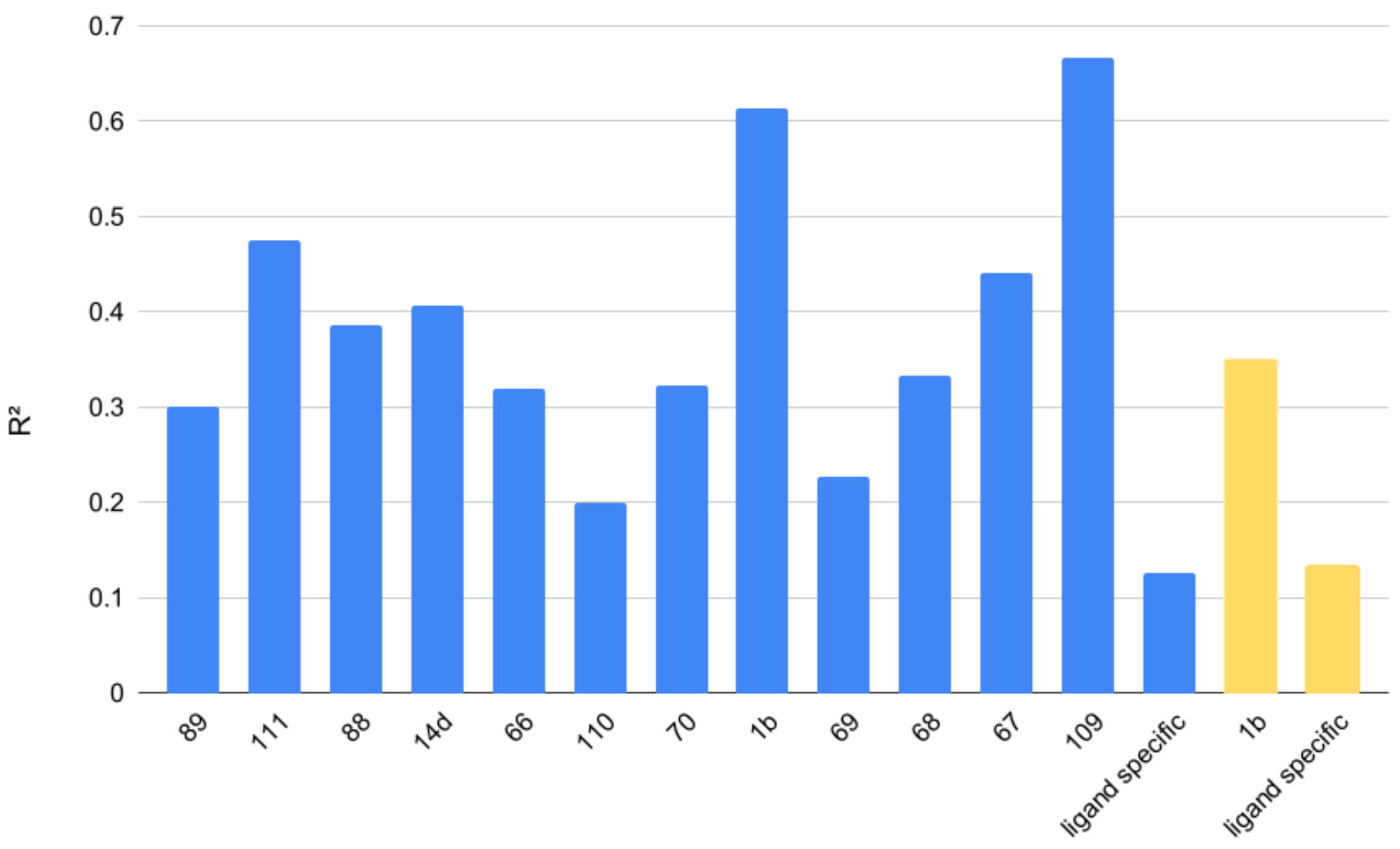}
    \caption{$R^2$ values for the full set of binding energies against pIC$_{50}$ for each set of protein charges. The blue and gold bars indicate RHF/STO-3G and $\omega$B97X/def2-TZVP levels of theory for the QM region, respectively.}
    \label{fig:spearmanrankings}
\end{figure}

We have investigated the effect of protein charges used to compute $E_{\text{ligand-in-protein (aq)}}$ on the ranking induced by computed binding energies and found that using ligand-specific charges (i.e. corresponding to the protein structure from each protein-ligand complex) yield much worse correlation than application of the same charges for all ligands, at both the RHF and DFT levels of theory, as seen on Figure \ref{fig:spearmanrankings}. The relatively large $R^2$ value obtained with charges from the 1b-BACE1 complex, along with the 1b ligand being the unsubstituted base-molecule of the oxazine series, motivated the choice of these charges for the quantum simulations.

\end{multicols}

\end{document}